\documentclass[runningheads]{llncs}
\usepackage[T1]{fontenc}
\usepackage[misc]{ifsym}
\usepackage{graphicx}
\begin{document}
\title{Contrastive Learning for Sleep Staging based on Inter Subject Correlation}
%
%
\author{Tongxu Zhang \and
Bei Wang\inst{(\textrm{\Letter})}}
\authorrunning{T. Zhang et al.}
%
\institute{School of Information Science and Engineering,
East China University of Science and Technology,
Shanghai, 200237, China\\
\email{juk@mail.ecust.edu.cn, beiwang@ecust.edu.cn}}
\maketitle              
\begin{abstract}
In recent years, multitudes of researches have applied deep learning to automatic sleep stage classification.
Whereas actually, these works have paid less attention to the issue of cross-subject in sleep staging.
At the same time, emerging neuroscience theories on inter-subject correlations can provide new insights for cross-subject analysis.
This paper presents the MViTime model that have been used in sleep staging study.
And we implement the inter-subject correlation theory through contrastive learning, providing a feasible solution to address the cross-subject problem in sleep stage classification.
Finally, experimental results and conclusions are presented, demonstrating that the developed method has achieved state-of-the-art performance on sleep staging.
The results of the ablation experiment also demonstrate the effectiveness of the cross-subject approach based on contrastive learning.
The code can be accessed through: \url{https://github.com/jukieCheung/MViTime}.
\end{abstract}
\section{Introduction}
Sleep is an essential part of everyone's life.
Good sleep is crucial in maintaining one's mental and physical health~\cite{siegel2005clues,maquet2001role}.
With the rapid development of modern society, high-intensity work and study,
irregular lifestyles affect the sleep of most people.
Currently, sleep problems have gradually become one of the important public health issues.

Sleep stage discrimination takes an important role to monitor sleep quality and diagnose sleep disorders.
Sleep stages are manually classified by physicians with professional qualifications and clinical experience for each sleep data segment.
Obviously, the manual classification of sleep stages is time-consuming.
Compared with manual classification, the algorithm/model-based automatic classification method is more efficient as an important tool in clinical applications.
In recent years, deep learning models have been applied to many fields.
With the increasing popularity of deep learning,
many automatic sleep stage classification methods based on convolutional neural networks (CNNs),
recurrent neural networks (RNNs), and transformers have been developed~\cite{supratak2017deepsleepnet,perslev2019u,phan2022sleeptransformer}.
Those deep learning models are all end-to-end networks, but have been only implemented using supervised learning.

Apart from this, there are still many problems that need to be investigated and explored in the practical clinical application of deep learning models for sleep stage classification.
The electrophysiological signals of different subjects inevitably have individual differences,
which will affect the actual discrimination performance of the well-trained model.
Studies in the literature~\cite{li2010application} and~\cite{buttfield2006towards} have demonstrated that EEG signals are unstable in cases of cross-subject due to user fatigue, different electrode placements, different impedances, etc.
Therefore, how to design and implement a sleep stage automatic classification model to deal with the problem of cross-subject is essential.

To solve the cross-subject problem,
researchers have applied domain adaptation (DA) methods on EEG signals~\cite{he2022single}.
However, domain adaptation methods require to access the test data, which cannot guarantee no information leakage problems.
At the same time, based on the emerging neuroscience theory of inter-subject correlation (ISC) and combined with contrastive learning methods,
a new approach can be provided to explore subject invariance and develop cross-subject analysis.
CLISA~\cite{shen2022contrastive} is such a work that starts from the perspective of inter-subject correlation and uses contrastive learning methods to perform emotional recognition based on EEG.
Contrastive learning has been widely used in many fields.
For example, in computer vision (CV), the representation of the network can be improved by learning from large amounts of unlabeled data based on the contrast between images~\cite{chen2020simple}.
In the existing literatures, only one study was found that used contrastive learning for automatic sleep staging~\cite{jiang2021self}.
However, the literature~\cite{jiang2021self} did not consider the inter-subject correlation but only compared the signals.

In summary, our study considers using comparative learning methods to distinguish sleep stages from the perspective of inter subject correlation,
in order to overcome the impact of individual differences. The ultimate purpose is to provide feasible and implementable solution for cross-subject sleep staging.
The main contributions of this paper is summarized as follows:

1. MobileViT is improved for time series tasks and combined with comparative learning methods to achieve optimal sleep stage classification.

2. Based on the method of comparative learning and the correlation between subjects, a cross-subject solution is designed for sleep staging.

\section{Method}
\subsection{Contrasting Learning Methods}
The contrastive learning approach used in this study is mainly based on the work of SimClr~\cite{chen2020simple}. However, the contrastive learning strategy is different including the self-contrast and cross-subject contrast which are designed to solve the cross-subject problem as well as to take into account the characteristics of the sleep EEG signal and the sleep staging task.

In practice, contrastive learning is a way of pre-training, where the input training set is used to initialise the parameters of the network by contrastive learning. The training set is re-input to fine-tune the network and train the classifier at the same time.

Applying contrastive learning to EEG signals requires learning the representation of EEG signals by calculating the similarity between different features in the cosine metric space. Here, note $x$ and $y$ as vector representations of the two EEG signals,
and note $sim(x, y)$ as the similarity measure of the two vectors $x$ and $y$,
\begin{equation}
sim(x, y)=\cos \left(\theta_{x, y}\right)=\frac{x^T y}{\|x\|\|y\|}, sim(x, y) \in[0,1]
\end{equation}
If the cosine value is close to 1, it indicates that the angle between the two vectors is smaller and the signals are more similar.
When the cosine value comes to 0, it indicates that the angle between the two vectors is larger and the signals are more different.

\begin{figure}
\centering
\includegraphics[width=0.4\textwidth]{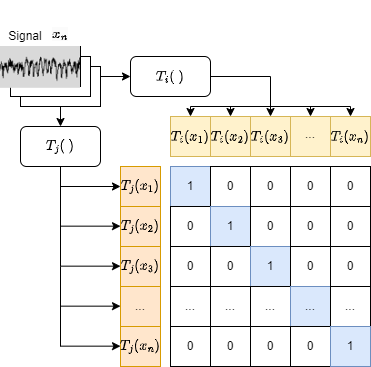}
\caption{Ground truth for contrastive learning of EEG signals.} \label{fig1}
\end{figure}

The self-contrast of the model is based on the work of SimClr,
where two different transformations of the signal are referred to the work of SSL~\cite{jiang2021self}.
The original signal sequence of $n$ epochs is firstly transformed twice respectively,
by Cropping and Permutation, to produce $2n$ transformed signals. It is then fed into the network to extract features.
During the training process, for each epoch, the similarity of the features is measured by the network.
In the remaining $2n-1$ samples, the corresponding identical signals but with different transformations are found, i.e. the network with the same epoch is taken as the positive pair and the rest as the negative pair.
The loss of the cross-entropy function is calculated using the ground truth as in Fig.~\ref{fig1}.

There are totally five sleep stages need to be identified, i.e. Wake, REM in the rapid eye movement phase, and S1, S2 and S3 in the non-rapid eye movement phase.
In order to implement the cross-subject contrast learning model, five epochs (each 30-second) corresponding to the above sleep stages are selected.
By using PCA (principle component analysis) for dimensionality reduction, a segment of 30 seconds is used as a feature for one subject.
Assuming that there are $n$ subjects, there are $n$ subject features.
Again, two different transformations are performed to generate $2n$ transformed signals that are fed into the network for feature extraction.
Different transformations of the same subjects are positive pairs and transformed signals between different subjects are negative pairs.
The same cross-entropy function is used to calculate the loss.

In addition, the above two different transformations are specified as,

Cropping: the raw EEG signal $T\left(t\right)$ ($\ t=1,2,\ldots,L$) is randomly separated into a number of segments $\left\{T_1\left(t\right),T_2\left(t\right),\ldots,T_n\left(t\right)\right\}$; a $T_r\left(t\right)$ is then selected randomly and is adjusted to the original length $L$.

Permutation: the raw EEG signal $T\left(t\right)$ ($\ t=1,2,\ldots,L$) is randomly separated into a number of segments $\left\{T_1\left(t\right),T_2\left(t\right),\ldots,T_n\left(t\right)\right\}$; those segments are disorganized into $\left\{T_{k_1}\left(t\right),T_{k_2}\left(t\right),\ldots,T_{k_n}\left(t\right)\right\}$ where $ \left\{k_1,\ k_2,\ldots,k_n\right\} $ is of $\left\{1,2,\ldots,n\right\}$; those segments are then reorganized together.

The above contrastive learning strategy can maximize the difference of a sample from other samples. Meanwhile, it considered the the inter-subject correlation to maximize the relevant data. 
Thus, for cross-subject sleep staging, the trained model can be applied to new subjects. 
It can improve the utility of automatic sleep stage classification based on deep learning models.

\subsection{Sleep Staging Contrastive Learning Model}
\begin{figure}
\centering
\includegraphics[width=\textwidth]{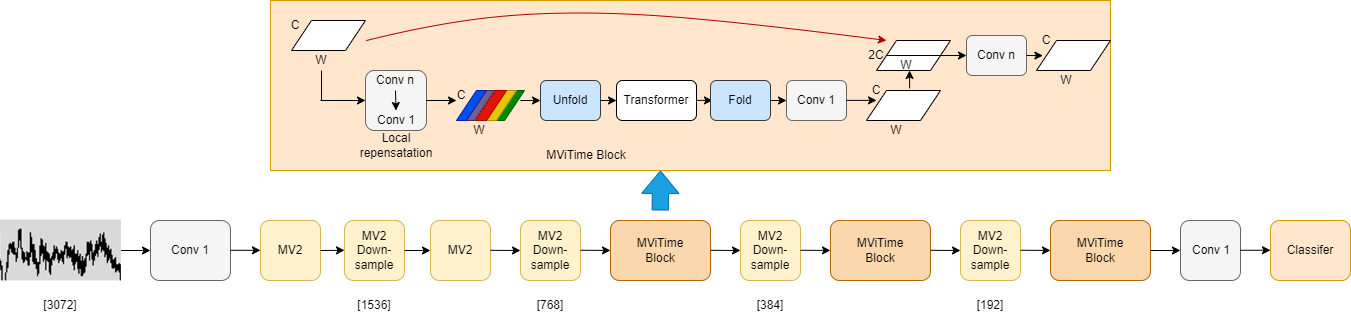}
\caption{Network structure of MViTime.} \label{fig2}
\end{figure}

The sleep stage classification model used in this study was mainly referred to the literature~\cite{hannun2019cardiologist} and~\cite{ronneberger2015u}.
The MobileViT from the literature~\cite{mehta2021mobilevit} was adopted due to the ability of transformer to extract long local features and the ability of CNN to extract global features.
The main difference is that MobileViT is an image model.
Therefore, we adjusted MobileViT to make it suitable for deep learning models dealing with one-dimensional time series. The developed model is named MViTime. The architecture of MViTime is shown in Fig.~\ref{fig2}.

\begin{figure}
    \centering
    \includegraphics[width=0.8\textwidth]{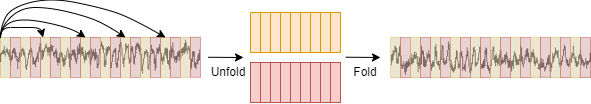}
    \caption{Unfold and Fold for MViTime.} \label{fig3}
\end{figure}

In Fig.~\ref{fig3}, the signal input to the MViTime module is divided into many segments of Tokens by Patch size. A position encoding is performed to represent the  characteristic of the signal.
Therefore, each Hertzian signal is related to the other signals in the MViTime module.
The tokens with the same colour are then rearranged and passed through the transformer for self-attention.
In this way, the amount of computation required for the self-attentive mechanism is reduced.

SiLU is used as the activation function and the optimizer is SGD,
where the SiLU function is smoother and less monotonic than the ReLU function,
which improves the generalisation performance of the model.
The batchsize of contrastive learning module is set to 128, mainly to allow the model to be loaded onto the GPU (Nvidia RTX 3090).
For the training of the classifier, the batchsize is set to 512, using a cosine warm-up strategy.

\subsection{Sleep Staging Classification Model}
\begin{figure}
    \centering
    \includegraphics[width=\textwidth]{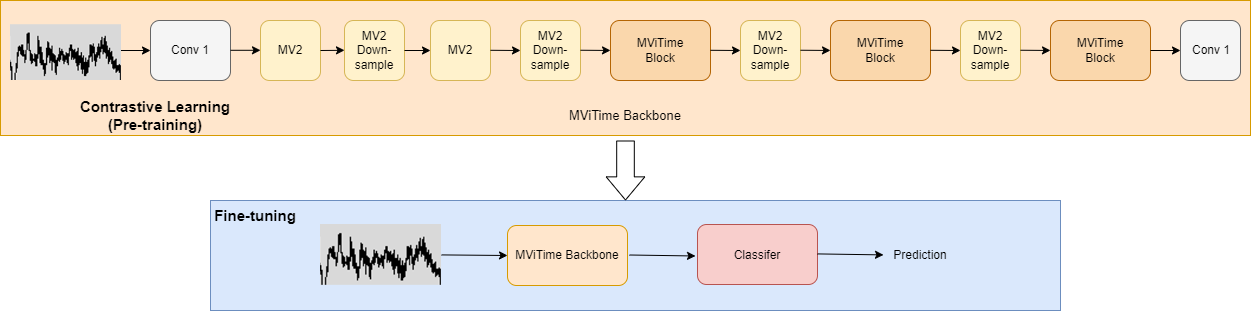}
    \caption{MViTime initializes the network parameters by pre-training with a contrastive learning method and subsequently fine-tunes these parameters.} \label{fig4}
\end{figure}

Fig.~\ref{fig4} shows the contrastive learning strategy for the developed MViTime model.
The input datasets are entered into the MViTime backbone network for pre-training to initialize the parameters by the contrastive learning method.
These datasets are re-entered to perform fine-tuning and train the classifier.
The reason for choosing fine-tuning instead of linear probe is that linear probe freezes the backbone to train the classifier,
i.e. it is equivalent to fine-tuning the last classification layer.
The neural network is supposed to be optimised as a whole to achieve the final classification. It is unlikely that only the last layer is operated.
Therefore, it is argued that overall fine-tuning is consistent with the mode of operation of neural networks.

\subsection{Cross-subject Strategy for Sleep Staging Based on Contrastive Learning}
\begin{figure}
    \centering
    \includegraphics[width=0.8\textwidth]{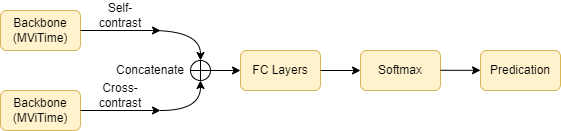}
    \caption{Cross-subject contrastive learning strategy for MViTime.} \label{fig5}
\end{figure}

Fig.~\ref{fig5} shows the cross-subject contrastive learning strategy for the MViTime model.
In the experiments of the cross-subject strategy, the output of backbone based on the self-contrast and the cross-subject contrast was weighted and combined. It is then loaded into the network to train the classifier.
The main objective was to maximise the difference between the different sleep stages and to maximise the difference between subjects.
There are two methods of combination:
one is that the contrastive learning modules of the two contrasts are combined but not the classifiers after the two contrasts;
another is that the whole network of the two contrasts are combined containing the contrastive learning modules and the classifiers.
Both methods are also trained using fine-tuning.

\section{Experiments}
\subsection{Dataset}
The polysomnographic data in this study were from the Sleep-EDF dataset~\cite{kemp2000analysis} , which contains 197 full-night polysomnographic recordings. The sleep staging markers in the Sleep-EDF dataset are based on the definitions and characteristics of sleep staging corresponding to the R$\&$K sleep staging guidelines. The Sleep-EDF data is divided into two parts, Sleep Cassette and Sleep Telemetry. There are totally 153 SC recordings (SC=Sleep Cassette) and 44 ST recordings (ST=Sleep Telemetry).

The Sleep-EDF dataset is the publicly available dataset,
of which the SC data is also the more commonly used dataset in research work on automatic sleep staging methods.
The amount of data used in the literature work varies depending on the processing method:
222,479 record data segments were used in the literature~\cite{mousavi2019sleepeegnet},
177,411 record data segments in the literature~\cite{he2022single} and 195,479 record data segments in the literature~\cite{eldele2021attention}.
In addition, the EEG signal in Sleep-EDF contained both Fpz-Cz and Pz-Oz channels. In the literature~\cite{he2022single} it is argued that the Fpz-Cz channel is closer to the eye than the Pz-Oz channel, and that this channel would be better to capture the characteristics of the sleep state corresponding to the electrical activity of eye movements.

In this study, the same 195,479 recording segments as in the literature~\cite{eldele2021attention} were considered and the Fpz-Cz single channel was chosen for automatic sleep stage classification.

Data from 78 healthy subjects from the Sleep-EDF dataset were used for the experiments.
The 78 healthy subjects constitute a dataset that is often divided into two datasets,
EDF-20 (also known as Sleep-EDF-2013) and EDF-78 (also known as Sleep-EDF-2018), where EDF-20 is the first 20 subjects out of 78 subjects.
The distribution of epochs corresponding to the five stages in the two datasets are shown in Table~\ref{tab1}.

\begin{table}
\centering
\caption{The number and percentage of 30-Second epochs of each sleep stage in the two datasets.}\label{tab1}
\begin{tabular}{|l|l|l|l|l|l|l|}
\hline
Dataset & Total number & Wake & Stage1 & Stage2 & Stage3 & REM\\
\hline
Sleep-EDF-20 & 42308 & 8285 & 2804 & 17799 & 5703 & 7717 \\
 &  & (19.6$\%$) & (6.6$\%$) & (42.1$\%$) & (13.5$\%$) & (18.2$\%$)\\
Sleep-EDF-78 & 195479 & 65951 & 21522 & 69132 & 13039 & 25835\\
 &  & (33.7$\%$) & (11.0$\%$) & (35.4$\%$) & (6.7$\%$) & (13.2$\%$)\\
\hline
\end{tabular}
\end{table}

\subsection{Performance Metrics}
To evaluate the classification performance of the presented model, there indicators included Accuracy, Recall and F1-score are calculated based on confusion matrix. The description of the indicators are described in Table~\ref{tab0}. 

\begin{table}
  \centering
  \caption{Meaning of Accuracy, Recall and F1-score..}\label{tab0}
  \begin{tabular}{|l|l|}
  \hline
  Indicators & Description                                                                                                                                                                                                        \\ \hline
  Accuracy   & \begin{tabular}[c]{@{}l@{}}This measures the proportion of correctly \\ classified instances among all instances.\end{tabular}                                                                                     \\ \hline
  Recall     & \begin{tabular}[c]{@{}l@{}}This measures the proportion of positive \\ instances that are correctly identified by the model.\end{tabular}                                                                          \\ \hline
  F1-score   & \begin{tabular}[c]{@{}l@{}}This is the harmonic mean of precision and recall. \\ It provides a balance between precision and recall, \\ taking into account both false positives and false negatives.\end{tabular} \\ \hline
  \end{tabular}
\end{table}

\subsection{Results of Sleep Staging Experiment}
\begin{figure}
    \begin{minipage}[t]{0.5\linewidth}
        \centering
        \includegraphics[width=\textwidth]{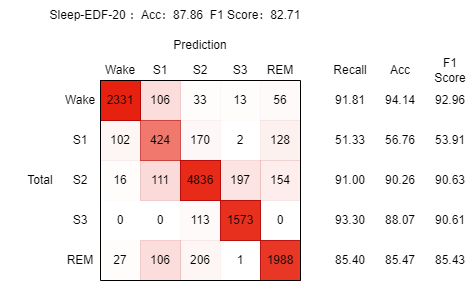}
        \centerline{(a) MViTime at Sleep-EDF-20}
    \end{minipage}%
    \begin{minipage}[t]{0.5\linewidth}
        \centering
        \includegraphics[width=\textwidth]{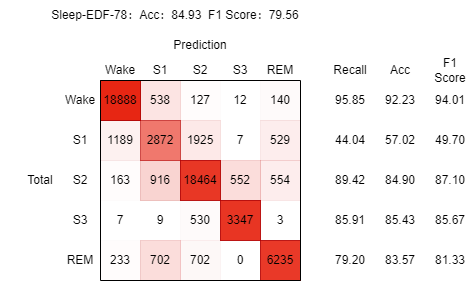}
        \centerline{(b) MViTime at Sleep-EDF-78}
    \end{minipage}
    \caption{Confusion matrix and performance metrics for MViTime} \label{fig6}
\end{figure}

Fig.~\ref{fig6} show the confusion matrix and performance metrics obtained by using our proposed method on the two datasets of Sleep-EDF-20 and Sleep-EDF-78, respectively. The row of confusion matrix represents the sleep staging inspected by physician, while the column represents the automatic classification results of sleep stages. The classification performance comparing with the visual inspection are evaluated by three indicators.

In addition, the obtained results by our method is compared with other deep learning methods in recent literatures.
The results of sleep staging is summarized in Table~\ref{tab2}, where the overall accuracy and F1-score, and the F1-score of each stage are given.  

\begin{table}
  \centering
  \caption{Performance comparison of the Sleep-EDF with state-of-the-art methods.}\label{tab2}
      \begin{tabular}{|l|l|l|l|lllll|}
      \hline
      Dataset & Method  & Acc & F1 & \multicolumn{5}{c|}{Per Class F1} \\
      \hline
                               &                          &                      &                     & W    & S1   & S2   & S3   & REM  \\
                               \hline
                               EDF-20  & DeepSleepNet~\cite{supratak2017deepsleepnet}     & 81.9                 & 76.6                & 86.7 & 45.5 & 85.1 & 83.3 & 82.6 \\
                               & SleepEEGNet~\cite{mousavi2019sleepeegnet}     & 81.5                 & 76.6                & 89.4 & 44.4 & 84.7 & 84.6 & 79.6 \\
                               & He's work~\cite{he2022single}       & 85.2                 & 78.9                & -    & -    & -    & -    & -    \\
                               & AttnSleep~\cite{eldele2021attention}       & 84.4                 & 78.1                & 89.7 & 42.6 & 88.8 & 90.2 & 79.0 \\
                               & XSleepNet~\cite{phan2021xsleepnet}      & 86.3                 & 80.6                & -    & -    & -    & -    & -    \\
                               & SSL~\cite{jiang2021self}              & 87.5                 & 82.1                & 92.4 & 51.5 & {\bfseries 90.9} & 90.6 & 85.1 \\
                               & MViTime(Our)             & {\bfseries 87.8}                 & {\bfseries 82.7}                & {\bfseries 93.0} & {\bfseries 53.9} & 90.6 & {\bfseries 90.6} & {\bfseries 85.4} \\
                               \hline
                               EDF-78                   & SleepEEGNet~\cite{mousavi2019sleepeegnet}      & 74.2                 & 69.9                & 89.8 & 42.1 & 75.2 & 70.4 & 70.6 \\
                               & U-Time~\cite{perslev2019u}           & 81.3                 & 76.3                & 92.0 & {\bfseries 51.0} & 83.5 & 74.6 & 80.2 \\
                               & He's work~\cite{he2022single}        & 83.9                 & 77.6                & -    & -    & -    & -    & -    \\
                               & AttnSleep~\cite{eldele2021attention}        & 81.3                 & 75.1                & 92.0 & 42.0 & 85.0 & 82.1 & 74.2 \\
                               & XSleepNet~\cite{phan2021xsleepnet}        & 84.0                 & 77.8                & -    & -    & -    & -    & -    \\
                               & SleepTransformer~\cite{phan2022sleeptransformer} & {\bfseries 84.9}                 & 78.9                & 93.5 & 48.5 & 86.5 & 80.9 & {\bfseries 84.6} \\
                               & SSL~\cite{jiang2021self}              & 84.2                 & 79.0                & 94.0 & 50.4 & 86.1 & 84.5 & 80.1 \\
                               & MViTime(Our)             & {\bfseries 84.9}                 & {\bfseries 79.6}                & {\bfseries 94.0} & 49.7 & {\bfseries 87.1} & {\bfseries 85.7} & 81.3\\
                               \hline
  \end{tabular}
\end{table}

The obtained results in Fig.~\ref{fig6} and Table~\ref{tab2} showed that our proposed method of MViTime achieved optimal overall performance metrics (State-of-the-art) on both the Sleep-EDF-20 and Sleep-EDF-78 datasets. According to the comprehensive indicator of F1-score of each sleep stage, the classification performance on sleep stages are rather satisfied especially of W, S2, S3 and REM. S1 is is a kind of transmission state from wake to sleep, the characteristic is difficult to precisely identified. Although the F1-score of S1 is lower than the other sleep stages, it still reached a fairly well level comparing with other deep learning models.

\subsection{Results of Cross-subject Experiments}
Further experiments are conducted to validate the cross-subject strategy based on MViTime.
Here, one subject is randomly selected from the Sleep-EDF-20 dataset. The selected sleep recording is used as the test data and the other subjects are as the training data.
The performance of MViTime and the effectiveness of cross-subject strategy based on MViTime are evaluated compared.
Table~\ref{tab3} showed the comparison results by using MViTime, MViTime+ and MViTime++ of three subjects. 
MViTime represents the developed deep learning model, MViTime+ and MViTime++ correspond to the two types of cross-subject strategy which is described in Section 2.4 and Fig.~\ref{fig5}.

\begin{table}
    \centering
    \caption{Effect of cross-subject protocol on five-stage F1 scores for three subjects.}\label{tab3}
    \begin{tabular}{|l|l|lllll|}
    \hline
    Subject                 & Method    & W    & S1   & S2   & S3   & REM  \\
    \hline
    SC4131 & MViTime   & 91.7 & 60.2 & {\bfseries 91.4} & {\bfseries 91.3} & {\bfseries 90.8} \\
                            & MViTime+  & {\bfseries 92.3} & {\bfseries 71.2} & 90.7 & 90.3 & 88.5 \\
                            & MViTime++ & 91.6 & 47.6 & 90.9 & 90.7 & 87.5 \\
    \hline
    SC4052 & MViTime   & 83.1 & 37.3 & 92.0 & 85.9 & 76.4 \\
                            & MViTime+  & 85.1 & 36.5 & 88.7 & 83.1 & 68.1 \\
                            & MViTime++ & {\bfseries 89.4} & {\bfseries 41.4} & {\bfseries 92.6} & {\bfseries 86.5} & {\bfseries 78.6} \\
    \hline
    SC4032 & MViTime   & 93.0 & 35.8 & 86.4 & 82.3 & 79.5 \\
                            & MViTime+  & 92.6 & 34.2 & 86.8 & 86.7 & 78.2 \\
                            & MViTime++ & {\bfseries 93.7} & {\bfseries 37.7} & {\bfseries 89.6} & {\bfseries 89.4} & {\bfseries 82.5} \\
    \hline
    \end{tabular}
\end{table}

In Table~\ref{tab3}, it can be observed that the cross-subject strategy of MViTime++ significantly improved the F1-score of S1 of subject SC4032 and SC4052E0, i.e. from 35.8$\%$ to 37.7$\%$ and from 37.3$\%$ to 41.4$\%$ respectively. For subject SC4131, MViTime and MViTime+ are slightly better than MViTime++. It seems that MViTime itself is already performing well enough which may prevent the cross-subject solution MViTime++ from optimising the network parameters further. Overall, however, the presented cross-subject strategy based on inter subject correlation has effectiveness on sleep staging when the deep learning model is facing to a new sleep recording with individual difference.

\subsection{Ablation}
Ablation experiments are conducted to verify the importance of contrastive learning pre-training and the potential for large-scale pre-training by removing the contrastive learning module.
The following models are defined:
(1) Baseline: MViTime network but without contrastive learning pre-training,
(2) Baseline + CL: MViTime network and with contrastive learning pre-training, which is the result in Section 3.3, 
(3) Baseline + CL-Large: MViTime network and with contrastive learning pre-training where the pre-training using the remaining 58 individuals in Sleep-EDF-78 by excluding the 20 subjects of Sleep-EDF-20.
The results of the ablation experiments are presented in Table~\ref{tab4}.

\begin{table}
    \centering
    \caption{Ablation experiments on the Sleep-EDF.}\label{tab4}
    \begin{tabular}{|l|l|l|l|lllll|}\hline
    Dataset & Method               & Acc & F1 & \multicolumn{5}{c|}{Per Class F1} \\
    \hline
                             &                                       &                      &                     & W    & S1   & S2   & S3   & REM  \\
                             \hline
                             EDF-20  & Baseline                              & 86.1                 & 80.0                & 92.5 & 46.0 & 89.6 & 89.1 & 82.9 \\
                             & Baseline + CL       & 87.8                 & 82.7                & 93.0 & 53.9 & 90.6 & 90.6 & 85.4 \\
                             & Baseline + CL-Large & 88.2                 & 82.1                & 93.2 & 48.7 & 90.9 & 91.0 & 86.5 \\
                             \hline
                             EDF-78  & Baseline                              & 83.6                 & 76.8                & 93.9 & 41.8 & 86.1 & 83.8 & 78.7 \\
                             & Baseline + CL       & 84.9                 & 79.6                & 94.0 & 49.7 & 87.1 & 85.7 & 81.3\\
    \hline
    \end{tabular}
\end{table}

It can be seen from Table~\ref{tab4} that 
the pre-training module of contrastive learning improves the classification performance of the model. Moreover,
in EDF-20, the accuracy of the model is improved from 87.8$\%$ to 88.2$\%$ when the data of 58 people unrelated to EDF-20 are used for the contrastive learning pre-training. The increasement in the amount of pre-training data leads to an improvement in the model effect.
The results show that the proposed contrastive learning module is effective in further improving the model performance and demonstrates the potential to bring improvements in model performance when the amount of data pre-trained for contrastive learning is increased.

In addition, the potential of large-scale contrastive learning pre-training on the cross-subject problem is explored.
The results are compared in Table~\ref{tab5}. Here, MViTime-58 is using MViTime without cross-subject strategy where contrastive learning pre-training using the remaining 58 individuals in Sleep-EDF-78. MViTime++ is the same as in Table~\ref{tab3}. 

\begin{table}
    \centering
    \caption{Effect of increased pre-training data on five stage F1 scores for three subjects.}\label{tab5}
    \begin{tabular}{|l|l|lllll|}\hline
    Subject                 & Method    & W    & S1   & S2   & S3   & REM  \\
    \hline
    SC4131 & MViTime-58 & 93.5 & 64.2 & 91.1 & 90.2 & 88.0 \\
                            & MViTime++ & 91.6 & 47.6 & 90.9 & 90.7 & 87.5 \\
                            \hline
    SC4052 & MViTime-58 & 89.9 & 47.3 & 94.5 & 93.4 & 76.9 \\
                            & MViTime++ & 89.4 & 41.4 & 92.6 & 86.5 & 78.6 \\
                            \hline
    SC4032 & MViTime-58 & 96.7 & 53.9 & 91.3 & 90.9 & 88.5 \\
                            & MViTime++ & 93.7 & 37.7 & 89.6 & 89.4 & 82.5 \\
    \hline
    \end{tabular}
\end{table}

From Table~\ref{tab5}, it can be seen that contrastive learning with an increased amount of pre-training data lead to an improvement in the model's performance across the subjects. MViTime-58 achieves significant performance of more than 5$\%$ across the three subjects in both Wake and the more difficult S1 which is comparable to or better than MViTime++ in S2, S3 and REM.
It is worth noting that MViTime-58 does not require additional inter-subject correlation contrasts compared to MViTime++,
but only an increase in the amount of pre-training data for self-contrast.
Such result is due to the internal structure of MViTime, as it has the self-attention mechanism of transformer.

\section{Conclusion}
In this study, a contrastive learning method is developed for sleep staging which is combined with the theory of inter-subject correlation. A deep learning sleep stage classification model of MViTime is proposed dealing with single-channel EEG signals. The sleep recording of two datasets from Sleep-EDF are utilized for evaluation. The main improvements of our method are validated and analyzed by several comparison experiments. The ablation experiments show that the contrastive learning module was effective and the importance of contrastive learning pre-training. Besides, the lightweight cross-subject strategy is more meaningful than large-scale pre-training in clinical application. Whereas, using contrastive learning for pre-training have significantly potential to improve classification performance and cross-subject performance when fine-tuning.

\subsubsection{Acknowledgements} This research was supported by the National Natural Science Foundation of China under Grant 61773164.

%
%
%

\begin{thebibliography}{8}
  \bibitem{buttfield2006towards}
  Buttfield, A., Ferrez, P.W., Millan, J.R.: Towards a robust bci: error
    potentials and online learning. IEEE Transactions on Neural Systems and
    Rehabilitation Engineering  \textbf{14}(2),  164--168 (2006)
  
  \bibitem{chen2020simple}
  Chen, T., Kornblith, S., Norouzi, M., Hinton, G.: A simple framework for
    contrastive learning of visual representations. In: International conference
    on machine learning. pp. 1597--1607. PMLR (2020)
  
  \bibitem{eldele2021attention}
  Eldele, E., Chen, Z., Liu, C., Wu, M., Kwoh, C.K., Li, X., Guan, C.: An
    attention-based deep learning approach for sleep stage classification with
    single-channel eeg. IEEE Transactions on Neural Systems and Rehabilitation
    Engineering  \textbf{29},  809--818 (2021)
  
  \bibitem{hannun2019cardiologist}
  Hannun, A.Y., Rajpurkar, P., Haghpanahi, M., Tison, G.H., Bourn, C., Turakhia,
    M.P., Ng, A.Y.: Cardiologist-level arrhythmia detection and classification in
    ambulatory electrocardiograms using a deep neural network. Nature medicine
    \textbf{25}(1),  65--69 (2019)
  
  \bibitem{he2022single}
  He, Z., Du, L., Wang, P., Xia, P., Liu, Z., Song, Y., Chen, X., Fang, Z.:
    Single-channel eeg sleep staging based on data augmentation and cross-subject
    discrepancy alleviation. Computers in Biology and Medicine  \textbf{149},
    106044 (2022)
  
  \bibitem{jiang2021self}
  Jiang, X., Zhao, J., Du, B., Yuan, Z.: Self-supervised contrastive learning for
    eeg-based sleep staging. In: 2021 International Joint Conference on Neural
    Networks (IJCNN). pp.~1--8. IEEE (2021)
  
  \bibitem{kemp2000analysis}
  Kemp, B., Zwinderman, A.H., Tuk, B., Kamphuisen, H.A., Oberye, J.J.: Analysis
    of a sleep-dependent neuronal feedback loop: the slow-wave microcontinuity of
    the eeg. IEEE Transactions on Biomedical Engineering  \textbf{47}(9),
    1185--1194 (2000)
  
  \bibitem{li2010application}
  Li, Y., Kambara, H., Koike, Y., Sugiyama, M.: Application of covariate shift
    adaptation techniques in brain--computer interfaces. IEEE Transactions on
    Biomedical Engineering  \textbf{57}(6),  1318--1324 (2010)
  
  \bibitem{maquet2001role}
  Maquet, P.: The role of sleep in learning and memory. science
    \textbf{294}(5544),  1048--1052 (2001)
  
  \bibitem{mehta2021mobilevit}
  Mehta, S., Rastegari, M.: Mobilevit: light-weight, general-purpose, and
    mobile-friendly vision transformer. arXiv preprint arXiv:2110.02178  (2021)
  
  \bibitem{mousavi2019sleepeegnet}
  Mousavi, S., Afghah, F., Acharya, U.R.: Sleepeegnet: Automated sleep stage
    scoring with sequence to sequence deep learning approach. PloS one
    \textbf{14}(5),  e0216456 (2019)
  
  \bibitem{perslev2019u}
  Perslev, M., Jensen, M., Darkner, S., Jennum, P.J., Igel, C.: U-time: A fully
    convolutional network for time series segmentation applied to sleep staging.
    Advances in Neural Information Processing Systems  \textbf{32} (2019)
  
  \bibitem{phan2022sleeptransformer}
  Phan, H., Mikkelsen, K., Ch{\'e}n, O.Y., Koch, P., Mertins, A., De~Vos, M.:
    Sleeptransformer: Automatic sleep staging with interpretability and
    uncertainty quantification. IEEE Transactions on Biomedical Engineering
    \textbf{69}(8),  2456--2467 (2022)
  
  \bibitem{ronneberger2015u}
  Ronneberger, O., Fischer, P., Brox, T.: U-net: Convolutional networks for
    biomedical image segmentation. In: Medical Image Computing and
    Computer-Assisted Intervention--MICCAI 2015: 18th International Conference,
    Munich, Germany, October 5-9, 2015, Proceedings, Part III 18. pp. 234--241.
    Springer (2015)
  
  \bibitem{shen2022contrastive}
  Shen, X., Liu, X., Hu, X., Zhang, D., Song, S.: Contrastive learning of
    subject-invariant eeg representations for cross-subject emotion recognition.
    IEEE Transactions on Affective Computing  (2022)
  
  \bibitem{siegel2005clues}
  Siegel, J.M.: Clues to the functions of mammalian sleep. Nature
    \textbf{437}(7063),  1264--1271 (2005)
  
  \bibitem{supratak2017deepsleepnet}
  Supratak, A., Dong, H., Wu, C., Guo, Y.: Deepsleepnet: A model for automatic
    sleep stage scoring based on raw single-channel eeg. IEEE Transactions on
    Neural Systems and Rehabilitation Engineering  \textbf{25}(11),  1998--2008
    (2017)

  \bibitem{phan2021xsleepnet}
  Phan, H., Ch{\'e}n, O.Y., Tran, M.C., Koch, P., Mertins, A., De~Vos, M.:
    Xsleepnet: Multi-view sequential model for automatic sleep staging. IEEE
    Transactions on Pattern Analysis and Machine Intelligence  \textbf{44}(9),
    5903--5915 (2021)

\end{thebibliography}
%

\end{document}